\documentclass[%
superscriptaddress,
 amsmath,amssymb,
 aps,
]{revtex4-1}

\usepackage{graphicx}
\usepackage{dcolumn}
\usepackage{bm}

\begin{document}

\preprint{APS/123-QED}

\title{Reply to the Comments on ``Curvature capillary migration of microspheres" by P. Galatola and A. W\"urger}

\author{Nima Sharifi-Mood}
\author{Iris B. Liu}%
\author{Kathleen~J.~Stebe}
 \email{kstebe@seas.upenn.edu}
\affiliation{Department of Chemical and Biomolecular Engineering, University of Pennsylvania, Philadelphia, PA, 19104, USA\\}%


\begin{abstract}
We have studied microparticle migration on curved fluid interfaces in experiment and derived an expression for the associated capillary energy $E$ for two cases, i.e., pinned contact lines \cite{disk} and equilibrium contact lines \cite{Sphere}, which differ from expressions derived by others in the literature. In this problem, a particle of radius $a$ makes a disturbance in a large domain characterized by principal radii of curvature $R_1$ and $R_2$. Since $a$ is smaller than all associated geometric and physico-chemical length scales, analysis calls for a singular perturbation approach.  We recapitulate these concepts, identify conceptual errors in the Comments about our work, and provide evidence from experiment and simulation that supports our view.
\end{abstract}

\pacs{Valid PACS appear here}
 \footnotetext{\textit{$^{a}$~Chemical and Biomolecular Engineering, University of Pennsylvania, Philadelphia, PA, 19104; E-mail: kstebe@seas.upenn.edu}}

\maketitle
We have studied microparticle migration on curved interfaces in experiment and derived expressions for the capillary energy $E$ for particles of radius $a$ for pinned contact lines \cite{disk} ($E_{\rm{PIN}}$) and equilibrium contact lines ($E_{\rm{EQ}}$) \cite{Sphere}.  Our expression for $E_{\rm{EQ}}$ differs from that derived by \citealt{Wurger}. We have performed experiments in which spheres \cite{Sphere} and disks \cite{disk} migrate in agreement with $E_{\rm{PIN}}$. Two Comments have been published challenging our conclusions \cite{Galatola,Wurger_C} for both contact line conditions. Below, we amplify arguments originally presented in our work \cite{disk,Sphere} to address conceptual errors in the Comments. We then address each objection raised concerning our work, with new evidence from experiment. \\
\\
\indent We assume that $ac\ll 1$, where $c$ is the largest principle curvature, so $aH_0\ll 1$ and  $a\Delta c_0\ll 1$, where $H_0$ is the mean curvature and $\Delta c_0$ is the deviatoric curvature of the host interface $h_0$. We further assume that $a$ is small compared to all other geometric length scales (e.g., the size of the domain $\sim c^{-1}$), physico-chemical length scales (e.g., the capillary length) and that the particle is far from boundaries. Both $h_0$ and the interface profile in the presence of the particle $h=h_0+\eta$ are found assuming $\nabla {h_0} \cdot \nabla {h_0} \ll 1$ and  $\nabla {h} \cdot \nabla {h} \ll 1$.  The disturbance made by the particle in the interface $\eta$ depends on the local curvature field.  Since $\eta$ scales with $a$ and decays over distances similar to $a$, analysis requires a singular perturbation treatment (see, e.g., Appendix A of \citealt{Sphere}). In its most general form, ${h_0}$ can be expanded in powers of $ac$ and described in a local coordinate $(r,\phi)$ in an inner domain which, several radii from the origin, matches to the interface shape in the outer domain (Fig.~1(a)). The inner expansion to order $(ac)^2$ is: 
\begin{eqnarray}
{h_0} = \frac{{{r^2}{H_0}}}{2} + \frac{{\Delta {c_0}}}{4}{r^2}\cos 2\phi,\label{host}
\end{eqnarray}
where $H_0$ and $\Delta c_0$ are slowly varying fields in the outer coordinate evaluated at the particle center of mass. They emerge as coefficients in Eq.~\ref{host} from a matching procedure performed at each power in $ac$.  Eq.~\ref{host} is a non-uniformly valid description of the interface, valid only for $r \sim a$. Physically, this implies that the interface can locally be described as the sum of bowl-like and saddle-like terms. However, this simple description cannot hold over larger length scales where, e.g., details of the vessel shape play a role. We find $\eta$ by solving ${\nabla ^2}h = 0$ for equilibrium (EQ) or pinning (PIN) boundary conditions on the particle surface, and by requiring that $\eta$ tends to zero as $r \to \infty$.  This limit implies considering $\eta$ and Eq.~\ref{host} over distances large compared to $a$, but small compared to the outer length scale $c^{-1}$. We find  $\eta$ to be a decaying function of $r$ that depends on the outer variable only via coefficients evaluated at the particle center of mass: \\
\begin{eqnarray}
{\eta _{{\rm{PIN}}}} = \left( {{h_p} - \frac{{\Delta {c_0}{a^2}}}{4}} \right)\frac{{{a^2}}}{{{r^2}}}\cos 2\phi,\label{pin}
\end{eqnarray}
\begin{eqnarray}
{\eta _{{\rm{EQ}}}} = \left( {\frac{{\Delta {c_0}{r_0}^2}}{{12}}} \right)\frac{{{r_0}^2}}{{{r^2}}}\cos 2\phi,\label{eq}
\end{eqnarray}
where $h_p$ is the amplitude of the quadrupolar mode excited by a particle with a pinned contact line. Both Eqs.~\ref{pin} and \ref{eq} decay to zero in the inner region. No additional \emph{ad~hoc} arguments are introduced far from the particle. Nor would such arguments be appropriate in evaluating the associated excess area of the interface, which is finite only in the inner region where $\eta$ is finite. \\
\begin{figure*} 
\centering
\includegraphics[scale=0.5]{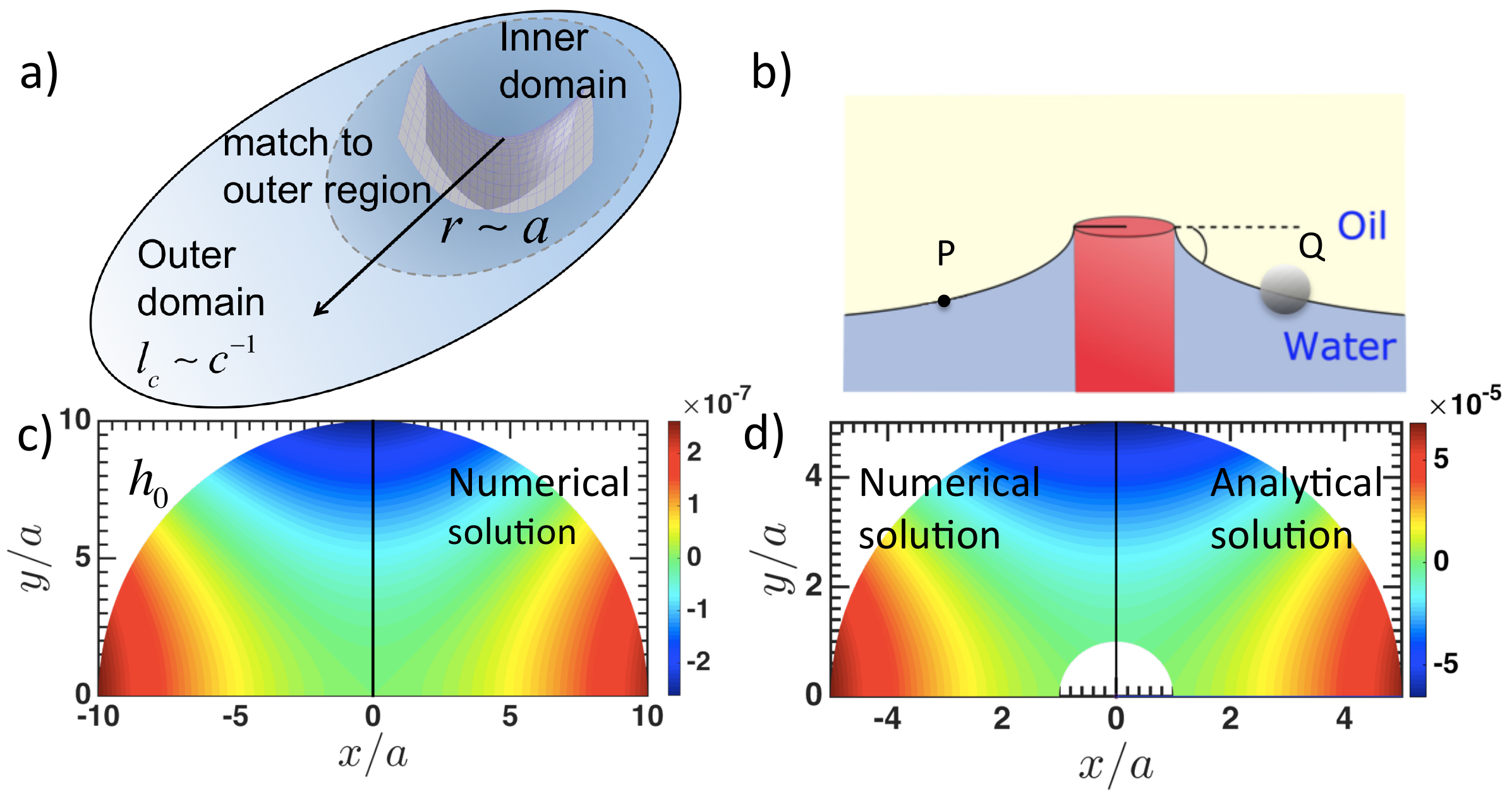}	     	    
 \caption{(a) Schematic of an inner domain which, several radii from the origin, matches to the interface shape of the outer domain. (b) Schematic of the interface shape simulated using boundary integral methods. An interface is pinned to a micropost of radius $R_m=125~\mu m$ with slope of 15$^\circ$--18$^\circ$. (c) Expanded view of the simulated host interface near $P$ in a plane tangent to the interface (right) and the analytical distortion field Eq.~\ref{host} (left). (d) Expanded view of the simulated interface near a sphere with equilibrium wetting conditions near $Q$ in a plane tangent to the particle center of mass (right), which agrees well with Eq.~\ref{eq} (left).}
\label{fig_1}
 \end{figure*}
 \\
\indent We have compared Eq.~\ref{host} to simulation (see, e.g., Fig.~4 in \citealt{Sphere}); we expand that discussion here. Using a boundary integral method, Laplace's equation is solved to find the interface shape in a setting similar to that in experiments (Fig.~1(b)), for which $H_0 \sim 0$. The host interface shape is determined, as is the shape of the interface in the presence of an attached particle. An expanded view of the simulated host interface (Fig.~1(c)) near a point $P$ far from boundaries agrees well with Eq.~\ref{host}. The quadrupolar mode of the simulated interface agrees with $h_0$ to better than $2\%$ for $r=10a$, with agreement improving closer to $P$. We also compare Eqs.~\ref{pin} and \ref{eq} to simulation near a point $Q$ far from boundaries, and find excellent agreement (see, e.g. Fig.~1(d) for $\eta_{\rm{EQ}}$). Finally, note that, for disturbance magnitudes corresponding to typical values from experiment, the distortions decay to negligible values, (sub-angstrom heights) within $r=10a$.\\
\\
\indent Returning to analysis, we find the associated energies.  
\begin{eqnarray}
{E_{{\rm{PIN}}}} = {E_0} - \gamma \pi {a^2}\left( {\frac{{{h_p}\Delta {c_0}}}{2} + \frac{{3{a^2}H_0^2}}{4}} \right),
\end{eqnarray}  
\begin{eqnarray}
{E_{{\rm{EQ}}}} = {E_p} + O\left( {{{\left( {a\Delta {c_0}} \right)}^4}} \right),\label{en-eq}
\end{eqnarray}  
where $E_0$ and $E_P$ are independent of position. In experiments with disks and microspheres on surfaces with $H_0 \sim 0$, we find $E$ to be consistent with $E_{\rm{PIN}}$; i.e., it varies linearly with $a\Delta c_0$, with the worst coefficient of linear regression $R^2=0.999$ for all cases. Eq.~\ref{en-eq} differs from prior work, which predicts a term for the curvature capillary energy that goes as ${(\Delta {c_0})^2}$. These differences arise owing to the treatment of a particular integral that appears in the excess area of the interface:
\begin{align}
&\int_{{r_0}}^\infty  {\int_0^{2\pi } {\left[ {\frac{{\partial {h_0}}}{{\partial r}}\frac{{\partial \eta }}{{\partial r}} + \frac{1}{{{r^2}}}\frac{{\partial {h_0}}}{{\partial \phi }}\frac{{\partial \eta }}{{\partial \phi }}} \right]r~d\phi dr} },\label{area_1}
\end{align}  
where $r_0 \sim a$ is the radius of the contact line on the particle. The discrepancies stem in taking $r \to \infty$ in evaluating this term. Before exploring this limit, we evaluate Eq.~\ref{area_1} for any finite slice between the contact line and a circle in the inner domain $r^{\ast}$. For equilibrium contact lines, this integral becomes:     
\begin{align}
&\int_{{r_0}}^{{r^*}} {\frac{{\Delta {c_0}^2}}{{12}}\frac{{{r_0}^4}}{r}dr\int_0^{2\pi } {( - {{\cos }^2}2\phi  + {{\sin }^2}2\phi )d\phi } }\nonumber\\  
&= \frac{{{r_0}^4\Delta {c_0}^2}}{{12}}\ln \left( \frac{{{r^*}}}{{{r_0}}}\right)(0) = 0.\label{area_2}
\end{align}  
The logarithmic term in Eq.~\ref{area_2} might cause concern, but the factor multiplying it is identically zero for any finite value of $r^{\ast}$.  To capture the value of the integral over the entire inner domain, we take the limit $r^{\ast}\to \infty$, recalling that, within the asymptotic framework, this implies a distance several particle radii away from the particle. The logarithmic factor then remains finite, and the integral is $0$.  This determines a pre-factor of $0$ for the terms quadratic in $\Delta c_0$ in $E_{\rm{EQ}}$ (similar arguments apply to the analysis for pinned contact lines).\\
\\
\indent In  \citealt{Wurger}, this integral is recast using the divergence theorem.  The area integral Eq.~\ref{area_1} can be written: 
\begin{align}
 \mathop {\lim }\limits_{r* \to \infty } \oint\limits_{r = r*} {{{\bf{e}}_r} \cdot (\eta \nabla {h_0}}) r~d\phi  - \oint\limits_{r = {r_0}} {{{\bf{e}}_r} \cdot (\eta \nabla {h_0}}) r~d\phi.\label{dd4}
\end{align} 
Equivalently, it can be written 
\begin{align}
 \mathop {\lim }\limits_{r^{\ast} \to \infty } \oint\limits_{r = r^{\ast}} {{{\bf{e}}_r} \cdot (h_0 \nabla {\eta}}) r~d\phi  - \oint\limits_{r = {r_0}} {{{\bf{e}}_r} \cdot (h_0 \nabla {\eta}}) r~d\phi.\label{dd5}
\end{align} 
In our analysis, the integrands in Eqs.~\ref{dd4} and \ref{dd5} are well defined quantities independent of $r$. Thus, in both Eqs.~\ref{dd4} and \ref{dd5}, the two contour integrals cancel, and, as required for equivalent representations of the same arguments, all three formulations agree. In W\"urger \cite{Wurger}, $\eta$ is assumed implicitly to decay to zero faster than ${1 \mathord{\left/{\vphantom {1 {{r^2}}}} \right.\kern-\nulldelimiterspace} {{r^2}}}$.  That is, while Eq.~\ref{eq} was presented as the local distortion around the particle, it was not used to evaluate the disturbance, except at the contact line. The contour integral at ${r^{\ast} \to \infty }$ is neglected in Eq.~\ref{dd4}, while that at the contact line is retained, yielding a finite term of order ${(\Delta {c_0})^2}$, reported as the primary result of his analysis (such an  operation would also yield a finite term of order ${(\Delta {c_0})^2}$ for the pinning conditions). However, the same reasoning would apply to the contour integral at ${r^{\ast} \to \infty }$ in Eq.~\ref{dd5}, yielding a finite contribution at the  inner contour which differs in sign from the result of Eq.~\ref{dd4}. Therefore, the result, in the absence of the outer contour, is not defined.  The disposal of the integral far from the particle is a failure to close the contours when applying the divergence theorem.\\
 \\
\indent In his Comment\cite{Galatola}, Galatola invokes a divergence of the slope of $h_0$ as ${r \to \infty }$.  To find a steep slope region in $h_0$, he takes $r \to c^{-1}$ in Eq.~\ref{host}. He recasts the contested area integral assuming large slopes, and argues that a modified form of the far field contour integral in Eq.~\ref{dd4} must be discarded. By this manipulation, he recovers W\"urger's result \cite{Wurger} for the equilibrium case. He applies this logic as well to the pinning case, altering coefficients and introducing a term quadratic in $\Delta c_0$.  Thus, he concludes that the proper curvature capillary energies should be:
\begin{eqnarray}
E_{{\rm{PIN}}}^{\rm{G}} = {E_0} - \gamma \pi {a^2}\left( {{h_p}\Delta {c_0} - \frac{1}{8}{a^2}\Delta {c_0}^2 - \frac{{{a^2}H_0^2}}{4}} \right),
\end {eqnarray}

\begin{eqnarray}
E_{\rm{EQ}}^{\rm{W}} = {E_p} - \frac{1}{{24}}{r_0}^2\Delta {c_0}^2,
\end {eqnarray}
where the superscripts ``W'' and ``G'' indicate W\"urger's and Galatola's arguments, respectively.  Intriguingly, in these suggested forms, the quadratic term in $E_{\rm{PIN}}^{\rm{G}}$ drives particles toward planar regions, while that in $E_{\rm{EQ}}^{\rm{W}}$ drives them to highly curved regions. \\ 
\\
\indent We counter argue that since Eq.~\ref{host} is not universally valid, it cannot be used to explore  $r \sim \ c^{-1}$.  In general, for interfaces obeying the assumptions under which Eq.~\ref{host} is derived, no steep slope regions are present. To support the assumption we make of small slopes made over the inner domain, we can estimate slopes of $h_0$ in Eq.~\ref{host} over the relevant range $a \le r \le {10a}$ using typical values from experiment, i.e., $a = 5 ~\mu$m, $6\times{10^{ - 3}} \le a\Delta {c_0} \le 1.2\times{10^{ - 2}}$ and $h_p\sim 25$~nm.  We find $\nabla {h_0} \cdot \nabla {h_0} \le 3.6\times{10^{ - 3}}$, and, hence, we remain justified in assuming $\nabla {h_0} \cdot \nabla {h_0} \ll 1$. A similar calculation shows $\nabla {h} \cdot \nabla {h} \ll 1$. \\
\\
 \indent Galatola further cites small departures of our data for $E$ versus $a\Delta c_0$ in the highest curvature regions of the data, which he states would be better fit by his theory. We noted above the linearity of the data when cast in terms of $E$ versus $a\Delta c_0$.  In Fig.~2(a), we reproduce the $E$ versus $a\Delta c_0$ data for a typical trajectory of a sphere; this was presented along with several others in \citealt{Sphere}.  We compare our theory to experiment.  In the inset, we also show contours for $E_{\rm{EQ}}^{\rm{W}}$ and $E_{\rm{PIN}}^{\rm{G}}$ for several values of $h_P$.  These predictions do not resemble the data. We derive a further test of functional form to compare to experiment. Recall that we study particle migration on interfaces formed around microposts of height $H_m$ and radius $R_m$ with interface slopes at the micropost's edge of $\psi \sim15^{\circ} - {18^{\circ}}$. The resulting interface shape has $H_0\sim0$ and $\Delta {c_0}(L) = {{2\tan \psi {R_m}} \mathord{\left/{\vphantom {{2\tan \psi {R_m}} {{L^2}}}} \right. \kern-\nulldelimiterspace} {{L^2}}}$, with corresponding capillary force according to our formulation of  ${F_{{\rm{curvature}}}} =  - {{d{E_{\rm{PIN}}}} \mathord{\left/
 {\vphantom {{d{E_{PIN}}} {dL}}} \right.
 \kern-\nulldelimiterspace} {dL}}\sim{{d\Delta {c_0}} \mathord{\left/
 {\vphantom {{d\Delta {c_0}} {dL}}} \right.
 \kern-\nulldelimiterspace} {dL}}\sim{1 \mathord{\left/
 {\vphantom {1 {{L^3}}}} \right.
 \kern-\nulldelimiterspace} {{L^3}}}$. In creeping flow, viscous drag ${F_{\rm{drag}}}$ balances ${F_{\rm{curvature}}}$. Noting that $F_{\rm{drag}}\sim {{dL} \mathord{\left/{\vphantom {{dL} {dt}}} \right.\kern-\nulldelimiterspace} {dt}}$ far from boundaries like walls and micropost, this implies that particles migrate with instantaneous position $L(t)$  which obeys a power law: 
\begin{eqnarray}
L(t) = {\left[ {B({t_f} - t) + L{{({t_f})}^4}} \right]^{1/4}},
\end{eqnarray}
where $B = {{2\pi {R_m}\tan \psi \gamma a{h_p}} \mathord{\left/{\vphantom {{2\pi {R_m}\tan \psi \gamma a{h_p}} {{C_D}\mu }}} \right.\kern-\nulldelimiterspace} {{C_D}\mu }}$ is constant, $C_D$ is the drag coefficient for an isolated particle, $L({t_f}) \sim 10a$ is the final position of the particle and $t_f$ is the time at which this position is attained. This power law is indeed obeyed (Fig. 2b inset); this agreement would not occur for the functional forms suggested by Galatola (see also Fig.~2(b), inset.) \\
\\
\indent Finally, we address the weak deviations from linearity noted by Galatola in $E$ vs.~$a\Delta c_0$ for particles  $ \sim 10a$ from contact with the micropost. Such deviations occur because of hydrodynamic and near field capillary interactions with the micropost. Of these, hydrodynamic interactions have the longest range. (In this calculation, since ${a \mathord{\left/{\vphantom {a {{R_{post}}}}} \right.\kern-\nulldelimiterspace} {{R_{\rm{post}}}}} = 0.04$, we neglect the curvature of  the post for particles close to contact.) We present a typical near-post trajectory in Fig. 2(c). Far from the micropost, the trajectory agrees with a trajectory predicted from $E_{PIN}$ and Stoke's drag on an isolated particle. Closer than $\sim 10a$ from contact, however, hydrodynamic interactions dominate. Using Brenner's drag formula \cite{Brennerwall} for a sphere near a wall and $E_{\rm{PIN}}$, predicted trajectories agree remarkably with experiment with no adjustable parameters.  This agreement fails very close to the wall, where the functional form of $h_0$ and $\eta$ must be amended owing to the presence of the wall. Note, however, for distances from the micropost between $ 10a$ and $5a$, the near-post particle trajectories are well described by our arguments for pinned contact lines when hydrodynamic interactions with boundaries are addressed.\\
\\
\indent In his comment, W\"urger considers a particle centered in a curved interface enclosed by a bounding circle at $R_{\rm{out}}$ assuming $a \ll R_{\rm{out}}$.  Imposing either pinned or equilibrium boundary conditions at $R_{\rm{out}}$, he retains reflected modes from the wall, with first contribution being of order ${({{{r_0}} \mathord{\left/{\vphantom {{{r_0}} {{R_{{\rm{out}}}}}}} \right.\kern-\nulldelimiterspace} {{R_{{\rm{out}}}}}})^4}$. He then argues that $\eta$ must be zero at $r=R_{\rm{out}}$, neglects the outer contour integral in Eq.~\ref{dd4}, and recovers  his prior result \cite{Wurger}. We counter that, in the asymptotic analysis, such reflected modes do not occur, as the distortion $\eta$ decays to zero for $r \ll R_{\rm{out}}$. Since $\eta$ is not finite near the wall, it is not influenced by the conditions there. The invocation of such boundary conditions on  $\eta$ at limits corresponding to vessel dimensions is a conceptual error within the framework of an asymptotic analysis. \\
\begin{figure*} 
\centering
\includegraphics[scale=0.7]{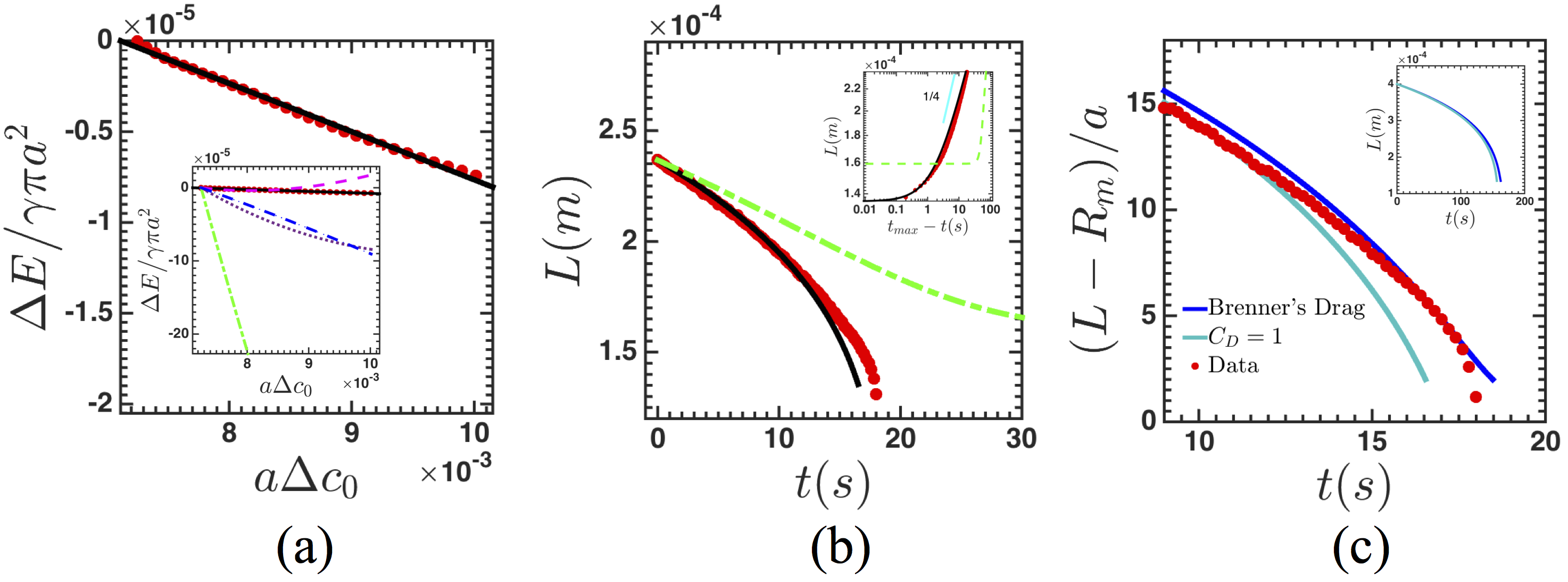}	     	    
 \caption{Curvature capillary energy of microspheres. (a) Comparison of predicted curvature capillary energy (solid black line) to that observed in experiment (red solid circle) for a single microsphere migrating near a post. Inset: Comparison of observed capillary energy with functional forms for $E_{\rm{PIN}}^{\rm{G}}$ for several values of $h_p$ (dashed pink $h_p=5$~nm, dotted violet $h_p=7$~nm and dashed green $h_p=20$~nm), and $E_{\rm{EQ}}^{\rm{W}}$ (dashed dark blue). (b) Comparison of experimental particle trajectory far from the micropost against predictions using Stoke's drag and $E_{\rm{PIN}}$ (solid black line), and $E_{\rm{PIN}}^{\rm{G}}$ (dashed green line). Inset: Particle trajectory shown in log-log plot showing agreement with power law predicted for $E_{\rm{PIN}}$. (c) Comparison of experimental particle trajectory against theory with Stoke's drag, i.e., $C_D=1$ and Brenner's drag for a spherical particle approaching a solid wall. Inset demonstrates the far field.} 
\label{fig_2}
 \end{figure*}
\indent While the geometry proposed by W\"urger does not correspond to that used in our experiments,  we can discuss the magnitude of the particle sourced disturbance and that of the suggested reflected modes for conditions corresponding to our experiments. We studied particles within $10a$ of contact with the micropost.  The coefficient of the particle-induced quadrupole in {$\eta_{\rm{EQ}}$}, ${{{r_0^2\Delta {c_0}} \mathord{\left/
 {\vphantom {{r_0^2\Delta {c_0}} {12}}} \right.
 \kern-\nulldelimiterspace} {12}}}$ is $\sim 5$~nm in our experiments.  This term decays as ${{{{a^2}} \mathord{\left/
 {\vphantom {{{a^2}} {{r^2}}}} \right.\kern-\nulldelimiterspace} {{r^2}}}}$. Thus, at $r=10a$, the disturbance would be $\sim 5\times10^{-11}$~m, a quantity too small to be considered finite in a continuum framework. If, however, this magnitude were deemed significant, one would generate reflected modes $\sim r^2$ whose magnitude near the particle would be 
$({{r_0^2\Delta {c_0}} \mathord{\left/
 {\vphantom {{r_0^2\Delta {c_0}} {12}}} \right.
 \kern-\nulldelimiterspace} {12}})({{r_0^4} \mathord{\left/
 {\vphantom {{r_0^4} {R_{{\rm{out}}}^4}}} \right.
 \kern-\nulldelimiterspace} {R_{{\rm{out}}}^4}})\sim5 \times {10^{ - 15}}$~m. The retention of terms of this magnitude, and their  integration over extended domains is not proper.\\ 
 \\
\indent To conclude, in this response, we show the Comments calling our results into question are based on a conceptual misunderstanding of the implications of limiting processes in singular perturbation analyses.  This is an interesting and potentially important example of effects owing to small inclusions in large domains, which have many physical manifestations in colloidal science.

\newpage
\acknowledgments
This work is partially supported by NSF grants CBET-$1133267$ and CBET-$1066284$, GAANN P200A120246, and MRSEC grant DMR11- $20901$.  We thank Nate Bade and Ningwei Li for discussions.  We gratefully acknowledge permission granted by Soft Matter to present material originally published in that journal.

%
\end{document}